\documentclass[10pt,twocolumn,aps,prl,showkeys]{revtex4-1}
\usepackage{amsmath}
\usepackage{amssymb}
\usepackage{graphicx}
\usepackage{xcolor}

\begin{document}

\title{Stability of a Rolled-Up Conformation State\\ for Two-Dimensional Materials in Aqueous Solutions}

\author{Maxim Trushin}
\affiliation{Centre for Advanced 2D Materials, National University of Singapore, Singapore 117546} 

\author{A. H. {Castro Neto}}
\affiliation{Centre for Advanced 2D Materials, National University of Singapore, Singapore 117546} 
\affiliation{Department of Material Science Engineering, National University of Singapore, Singapore 117575}

\date{\today}

\begin{abstract}
  Two-dimensional (2D) materials can roll up, forming stable scrolls under suitable conditions.
  However, the great diversity of materials and fabrication techniques has resulted in a huge parameter
  space significantly complicating the theoretical description of scrolls.
  In this Letter, we describe a universal binding energy of scrolls determined solely by their material parameters,
  the bending stiffness, and the Hamaker coefficient.
  Aiming to predict the stability of functionalized scrolls in water solutions,
  we consider the electrostatic double-layer repulsion force that may overcome the binding energy and
  flatten the scrolls.
  Our predictions are represented as comprehensive maps indicating the stable and unstable regions of a rolled-up conformation state
  in the space of material and external parameters.
  While focusing mostly on functionalized graphene in this work,
  our approach is applicable to the whole range of 2D materials able to form scrolls.
\end{abstract}


\maketitle

\begin{figure}[t]
  \includegraphics[width=0.8\columnwidth]{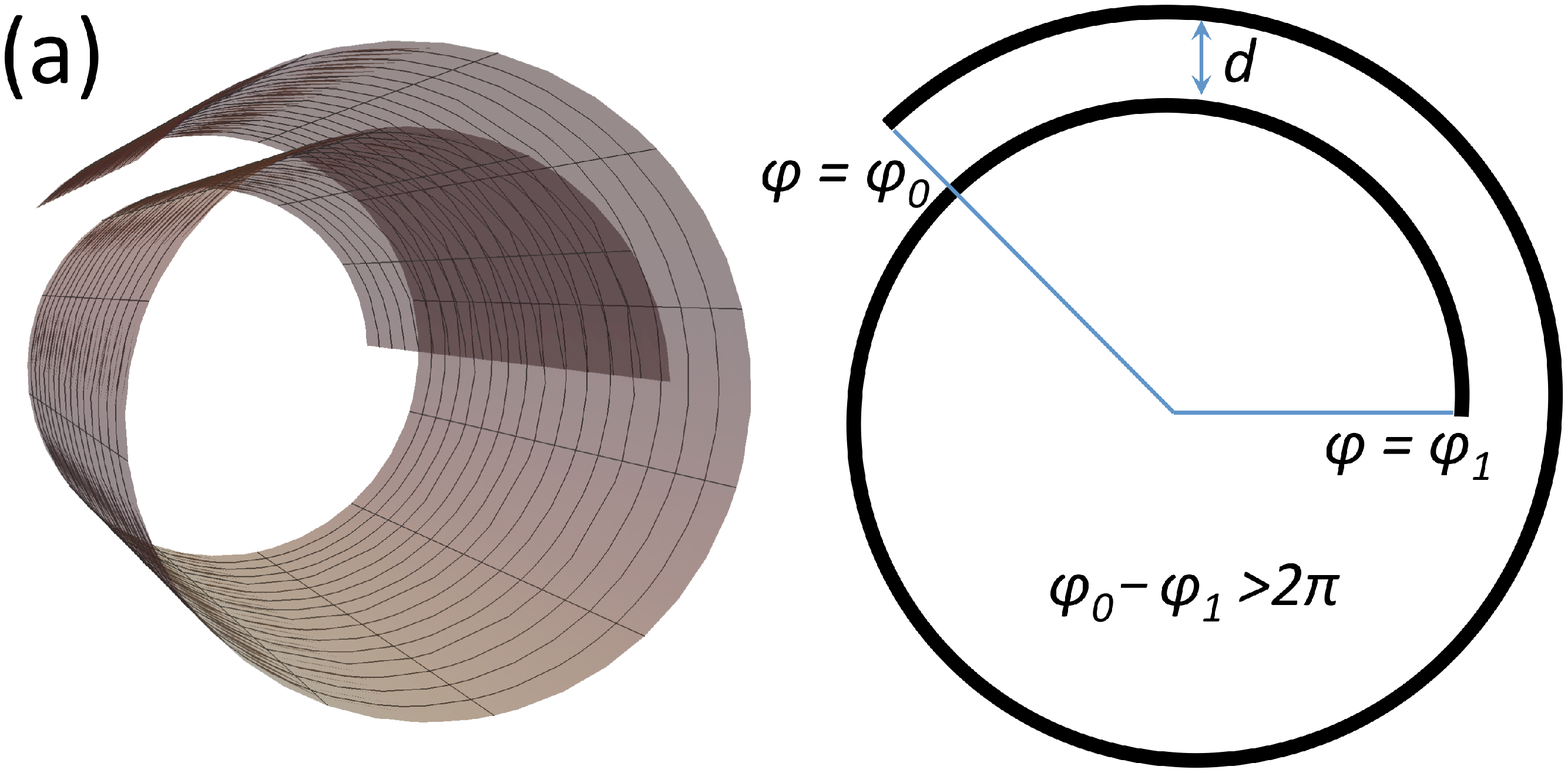}
  \includegraphics[width=0.8\columnwidth]{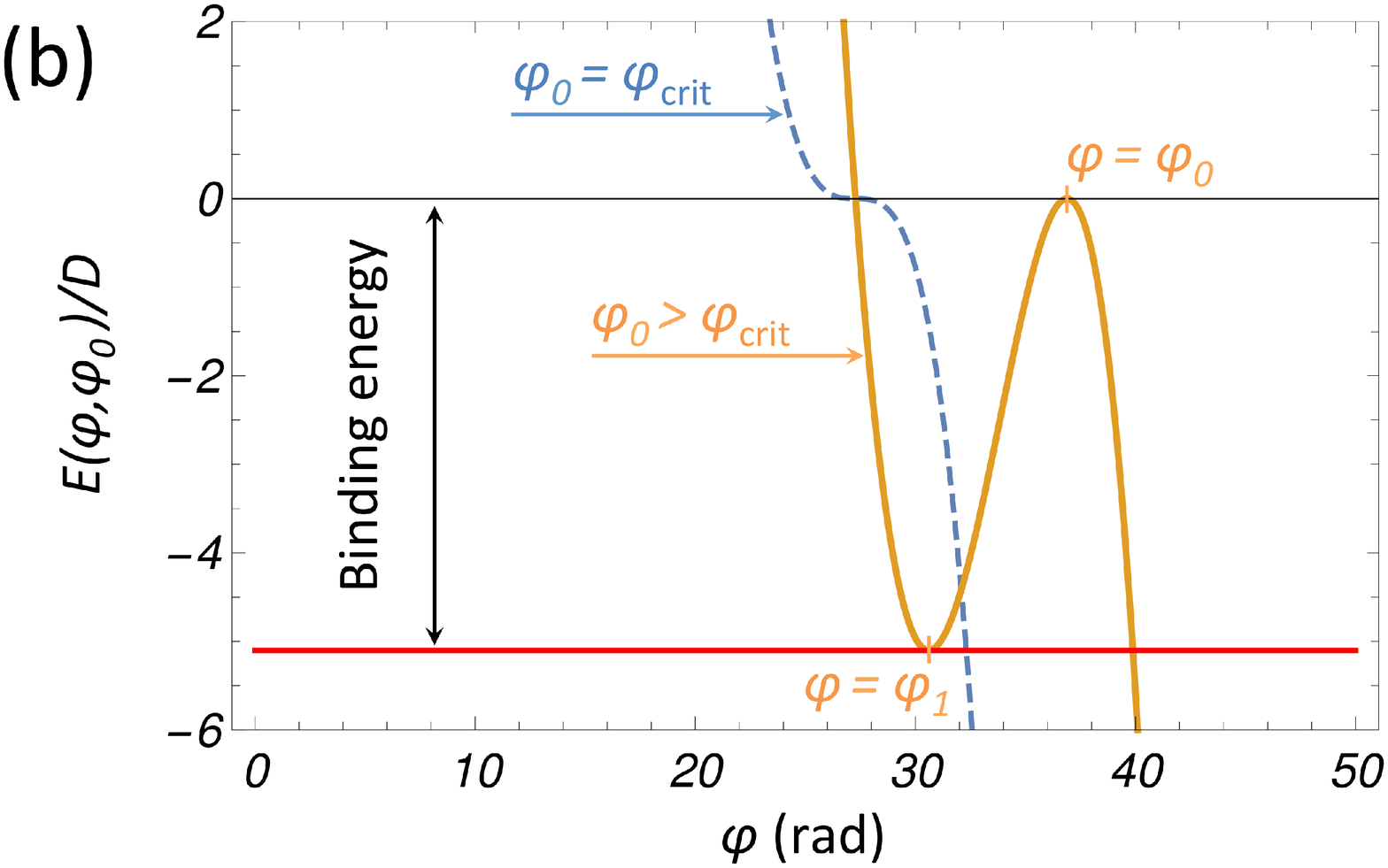}
  \caption{(a) Schematic image of a rolled-up 2D flake and the model parameters involved.
  The shape is given by Eq.~(\ref{shape}).
  (b) Energy of a scroll given by Eq. (\ref{energy}) at a fixed $\varphi_0$ as a function of $\varphi$ in the case 
  of equal bending stiffness and the Hamaker constant $D=H$.
  The energy behaves qualitatively similarly at any reasonable difference between $D$ and $H$.
  There is no local energy minimum (hence, no binding energy) at $\varphi_0=\varphi_\mathrm{crit}$, given by Eq.~(\ref{crit}).
  The binding energy exists at $\varphi_0>\varphi_\mathrm{crit}$. Here, $\varphi_{0,1}$ are given by Eq.~(\ref{phi01}).}
  \label{fig1}
\end{figure}

{\em Introduction. ---}
Rolling up a microscopic solid structure is a highly nontrivial way to alter its properties \cite{REVIEW2019,nature2021,prinz2000,schmidt2001nanotechnology},
e.g. its electron magnetotransport \cite{magnetotransport-2007schmidt}, plasmon \cite{plasmon-2009mendach}, and spin \cite{spin-2007trushin,spin-2010mendach} dynamics.
The recent advent of two-dimensional (2D) materials has widened the research field to include ultrathin scrolls \cite{graphene-scrolls-2009,CVDgraphene-scrolls-2014,MoS2-scrolls-2016}.
The functionalization ability of 2D scrolls \cite{graphene-functionalization-theory,graphene-functionalization-review,MoS2-functionalization,GO-functionalization} has ignited interest in various applications ranging from {\em superlubricants} for nanoparticles \cite{superlubricity-2015} to {\em supercapacitors}
for miniaturized electronics \cite{supercapacitors-2012,supercapacitors-poly-2019}.
Certain functionalized 2D materials allow for a transition between flat and rolled-up conformation states depending
on the solution content and can be seen as 2D polyelectrolytes \cite{ourAdvMat}.
This fact opens possibilities for the mass production of 2D scrolls in a solvent \cite{LPE2009coleman,graphene-scrolls-chem,scrolls-ultrasonic,GO-scrolls-lyophilization}.
Further development of fabrication techniques and potential applications of 2D scrolls 
require a theoretical model predicting the stability of the scrolls in aqueous solutions subject to external conditions.


The stability of 2D scrolls depends on multiple factors such as bending stiffness, thermal fluctuations, 
wrinkling \cite{strained-roll-up-theory2010}, edge defects \cite{defect-roll-up-theory2011}, and, above all, the interlayer adhesion.
Analytical models \cite{old-theory-review,Lennard-Jones-circles} and molecular dynamic simulations \cite{Lennard-Jones-circles,carbon-scrolls-mol-dyn,carbon-scrolls-mol-dyn2}
agree that the major contributions come from elastic and van der Waals forces playing competing roles.
However, once immersed in water, the functionalized scrolls experience additional double-layer electrostatic interactions that
can break the balance of forces and flatten the scrolls back to flakes \cite{GO-roll-up-theory2018}.
As the scrolls can be made of different materials with various functionalizations,
an explicitly solvable model is needed to describe all possible parameter combinations at once.
Despite recent theoretical efforts \cite{old-theory-review,Lennard-Jones-circles,GO-roll-up-theory2018} the solutions found
are still far from universal, limiting their applicability.
Here, we offer an elegant solution of the scroll stability problem, mapping all relevant interactions
onto the Archimedean spiral --- the most natural shape for any rolled-up elastic band.
The model applies to a broad range of 2D materials --- from superflexible \cite{GO-bending2016} graphene oxide (GO) to much more rigid \cite{graphene-bending-2009} graphene.

{\em Model. ---}
The scroll shape shown in Fig.~\ref{fig1}(a) 
can be described in polar coordinates $\{\varphi,r\}$ by the following simple equation:
\begin{equation}
\label{shape}
 r_\varphi  =  \frac{\varphi}{2\pi} d,
\end{equation}
where $d$ is the interlayer distance and its relation to the linear size $L$ is given by
\begin{eqnarray}
\label{archimed}
 L(\varphi_0,\varphi_1) & = & \int\limits_{\varphi_1}^{\varphi_0} d\varphi 
\sqrt{r_\varphi^2 + \left(\frac{dr_\varphi}{d\varphi}\right)^2} \\
\nonumber &=& \frac{d}{4\pi}\left(\varphi_0 \sqrt{1+\varphi_0^2} - \varphi_1\sqrt{1+\varphi_1^2} \right.\\
\nonumber &&\left. +\mathrm{arcsinh}\,\varphi_0 -
\mathrm{arcsinh}\,\varphi_1 \right).
\end{eqnarray}
The typical images of scrolls \cite{ourAdvMat,GO-roll-up-theory2018} indeed suggest uniform interlayer separation
consistent with our assumption.
Equation~(\ref{archimed}) makes sense for scrolls as long as $|\phi_0-\phi_1|\geq 2\pi$.
Note that the interlayer distance does not depend on the winding number even though the radius of curvature does.
In what follows, we find a universal ({\em i.e.,} independent of the geometrical parameters $d$ and $L$)
expression for the scroll binding energy and consider the stability of scrolls subject to  electrostatic double-layer repulsion. 

{\em Binding energy. ---} The binding energy of a scroll is determined by the interplay between the bending stiffness $D$ trying to unroll the scroll
and the counteracting van der Waals interlayer adhesion parameterized by the Hamaker constant $H$.
The elastic energy \cite{nelson2004statistical,RevModPhys2009} can be expressed through the surface integral
over the local curvature squared, $E_\mathrm{el} = \frac{D}{2}\int d^2\mathbf{r} R^{-2}(\mathbf{r})$, and the van der Waals energy 
is dominated by the intermolecular London contribution \cite{israelachvili2015intermolecular,ceramics-hamaker-2000}
given by $E_\mathrm{mol} =  - \frac{H}{12\pi d^2} A_\mathrm{o}$, with $A_\mathrm{o}$ being the interlayer overlap area.
We estimate the area as $A_\mathrm{o}\sim L^2$, with $L$ being the characteristic size of the flakes.
We map both energy contributions onto the Archimedean spiral with $\varphi_0$ fixed and
$\varphi$ relaxed [Fig. \ref{fig1}(a)].
We minimize the total energy $E(\varphi_0,\varphi)=E_\mathrm{el}+E_\mathrm{mol}$ and 
find $\varphi=\varphi_1$ at which the Archimedean scroll stabilizes [Fig. \ref{fig1}(b)].
The differential geometry suggests 
$1/R(\varphi)=|r_\varphi^2 + 2 r_\varphi^{'2} -r_\varphi r''_{\varphi\varphi}|/(r_\varphi^2 + r_\varphi^{'2})^{3/2}$;
hence, $E_\mathrm{el}$ reads
\begin{eqnarray}
\label{el}
 E_\mathrm{el} &  =  & \frac{DL}{2} \int\limits_{\varphi}^{\varphi_0}d\varphi\, r_\varphi / R^2(\varphi) \\
 \nonumber & = & \frac{\pi DL}{4d}\left[\frac{1}{(1+\varphi^2)^2}  + \frac{4}{1+\varphi^2} \right. \\
 \nonumber && \left. + 2 \ln \left(\frac{1+\varphi_0^2}{1+\varphi^2}\right) - \frac{5+4\varphi_0^2}{(1+\varphi_0^2)^2} \right].
\end{eqnarray}
{Note that both elastic $E_\mathrm{el}\propto L/d$ and adhesion $E_\mathrm{mol}\propto L^2/d^2$ energies
are scaled by the ratio $L/d$, which can be written in terms of $\varphi$ and $\varphi_0$ by means of Eq.~(\ref{archimed}) with $\varphi_1=\varphi$.
This allows us to get rid of the parameters $L$ and $d$ altogether and develop a universal model in terms of the angles $\varphi$ and $\varphi_0$.
The approach would not work for the non-Archimedean geometries.

Despite the relative simplicity of $E_\mathrm{mol}$ and $E_\mathrm{el}$, the resulting $E(\varphi_0,\varphi)$ 
is still difficult to analyze.
We therefore expand $E(\varphi_0,\varphi)$ in terms of 
$\Delta\varphi = \varphi - \varphi_0 \ll \varphi,\varphi_0$ and write $E(\varphi_0,\varphi)$ as 
\begin{equation}
 E(\varphi_0,\varphi) = D\frac{\Delta\varphi^2}{2} - H\frac{\varphi^2 \Delta\varphi^2}{48\pi^3}, \quad  \Delta \varphi > 2\pi.
 \label{energy}
\end{equation}
Note that $\Delta \varphi > 2\pi$ because 
the interlayer overlap area is finite if and only if the scroll makes at least one full turn.
The price we pay for the expansion in terms of $\Delta \varphi$ 
is that we can no longer find an exact value for $\Delta\varphi$, which should be just taken close to $2\pi$.
{Nonetheless, $E(\varphi_0,\varphi)$ correctly reproduces the physical picture:
(i) the energy rapidly increases up to a large positive value at small $\varphi$, indicating strong elastic strain,
and (ii) drops down to negative infinity $\propto -\varphi^4$ at $\varphi\to\infty$ describing the unrolling process.
There is also a cubic term $\propto 2\varphi_0\varphi^3$ that is responsible for a local energy minimum at which the scroll may stabilize.

Limiting ourselves to positive $\varphi_0$ and $\varphi$, we find 
that the local energy minimum  disappears at $\phi_0=\varphi_\mathrm{crit}$ given by
\begin{equation}
 \frac{\varphi_\mathrm{crit}}{2\pi} = \sqrt{6\pi\frac{D}{H}};
 \label{crit}
\end{equation}
see Fig. \ref{fig1}(b). To be specific, consider $\varphi_0 > \varphi_\mathrm{crit}$. 
In this case, the energy local maximum is at $\varphi=\varphi_0$, and the local minimum is shifted to the left from the maximum by
$\Delta\varphi=3\varphi_0/4 - \sqrt{\varphi_0^2/16 +\varphi_\mathrm{crit}^2/2}$; see Fig.~\ref{fig1}(b).
Since $\Delta\varphi$ must be small but not less than $2\pi$, we set the border value $\Delta\varphi=2\pi$, and
after some algebra we obtain the coordinates for both the local energy maximum ($\varphi_0$) and minimum ($\varphi_1$) as
\begin{equation}
 \frac{\varphi_{0,1}}{2\pi} = \left\{\begin{array}{c} 3/2 \\ 1/2 \end{array}\right\}
 + \sqrt{\left(\frac{\varphi_\mathrm{crit}}{2\pi}\right)^2 + \frac{1}{4}}.
 \label{phi01}
\end{equation}
The binding energy $E_b$ can be found easily from Eq.~(\ref{energy}) as $E_b=-E(\varphi_0,\varphi_1)$
with $\varphi_{0,1}$ given by Eq. (\ref{phi01}); see also the red line in Fig. \ref{fig1}(b).
The binding energy depends solely on the material parameters $H$ and $D$ no matter how large the scroll is. 
In the limiting case of either very soft \cite{GO-bending2016} or very stiff \cite{graphene-bending-2009} material,
$E_b$ can be written explicitly as
\begin{equation}
 E_b = \left\{
\begin{array}{ll}
\sqrt{2\pi^2 HD/3}, & H\ll D,\\
 2\pi^2 D + \pi H/3, & H\gg D.
 \end{array}
 \right.
\end{equation}
The limit $H/D\to \infty$ cannot be described as it implies $\varphi_\mathrm{crit}\to 0$, which contradicts our initial assumption
of large $\varphi_{0,1}$. Physically, the scrolls collapse, losing any regular structure in that limit.
The material may also be too rigid to form scrolls at $D \gg H$. 

\begin{figure}
  \includegraphics[width=\columnwidth]{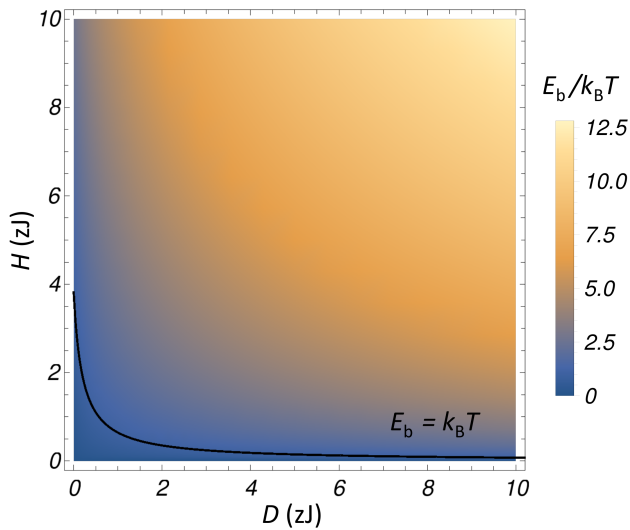}
  \caption{Scrolls' binding energy ($E_b$) vs Hamaker constant ($H$) and bending stiffness ($D$).
  The scrolls are unstable in the blueish region, where $E_b$ is of the order of $k_B T$ at room temperature $T$.
  The figure demonstrates that the values of {\em both} the Hamaker constant and the bending stiffness must not be smaller than $k_B T$
  to stabilize the scrolls.}
  \label{fig2}
\end{figure}

Figure~\ref{fig2} shows $E_b$ for arbitrary pairs of $H$ and $D$. 
The Hamaker constant may reach one hundred of zJ in vacuum \cite{israelachvili2015intermolecular,graphite-hamaker-2002},
but it is reduced at least by an order of magnitude in water \cite{israelachvili2015intermolecular,ceramics-hamaker-2000}.
The exact value strongly depends not only on the parent material but also on the layer thickness \cite{graphene-hamaker-2018},
layer curvature \cite{graphene-hamaker-comp-2007}, interlayer media \cite{GO-hamaker-on-sand}, and functionalization \cite{GO-Hamaker-comp-2016}.
The bending stiffness is about 100 zJ for pristine graphene  \cite{graphene-bending-2009},
but it drops drastically down to 4 zJ upon functionalization \cite{GO-bending2016}.
If either $D$ or $H$ is too low, then the binding energy is low and the scrolls are unstable in water.
To maintain the stability of the scrolls, we need not only good interlayer adhesion but also a suitable material elasticity.

Our model gives reasonable predictions for the actual GO scrolls studied in Ref. \cite{GO-roll-up-theory2018}.
The inner radius $r(\varphi_1)=2.5$ nm and interlayer distance $d=0.62$ nm results in 
$\phi_1\sim 25$ rad, very close to the model value at $H/D \sim 1$; see also Fig.~\ref{fig1}(b).
Taking $D=4$ zJ for GO \cite{GO-bending2016}, we obtain the reasonable $E_b\sim 5 k_B T$
suggesting that the scrolls are stable at room temperature $T$.
(Here, $k_B$ is the Boltzmann constant.) A similar analysis can be performed
for graphene with alternative functionalizations \cite{ourAdvMat}.



{\em Stability of scrolls in aqueous solutions. ---} Functionalized 2D materials immersed in an aqueous solution may acquire a surface charge that,
in turn, creates a certain electrostatic potential.
The resulting electrostatic repulsion between neighboring winds may unroll the scroll.
In an electrolyte, however, the counterions are able to screen the interlayer repulsion \cite{SM}.
The counterions form the Stern layer at the interface and the diffuse layer farther away from the charged surface.
The former contains immobile counterions and reduces the true surface potential down to the so-called $\zeta$ potential measured 
in the diffuse layer \cite{GO-functionalization}.
The diffuse layer contains mobile counterions, resulting in osmotic pressure \cite{israelachvili2015intermolecular}.

The physics behind the interlayer pressure is based on the solution of the Poisson-Boltzmann equation \cite{israelachvili2015intermolecular}.
The equation can also be solved in the cylindrical coordinates relevant for our geometry \cite{curvature1981,curvature1985,quasiplanar2000,quasiplanar2003}.
Although our surfaces are curved, the radius of curvature is much larger than the interlayer distance.
One can show that the solution converges to the well-known planar expression in this case \cite{SM}.

In contrast to the elastic and van der Waals interlayer adhesion energies,
the electrostatic double-layer repulsion energy does not scale by the ratio $L/d$,
spoiling the universality if our model.
We could have certainly considered all the interactions on equal footing,
minimizing the energy with respect to $\varphi$ in the same way as it has been done in Eq. (\ref{energy}).
However, such an approach would be physically incorrect.
The double-layer repulsion is supposed to unroll the scrolls. This is how we want to probe the stability of the rolled-up structure.
Upon unrolling the scroll, the scroll's geometry inevitably changes and can no longer be described by the Archimedean spiral 
invalidating our main assumption.
Instead of guessing the shape evolution upon unrolling, we  
follow the thermodynamic approach and compare the initial (rolled-up state) and final (unrolled state) energies, figuring out which is lower. 

To do that, we introduce the enthalpy difference between rolled-up and unrolled conformation states:
$\Delta {\cal H} = \Delta U + \Delta W$,
where the internal energy difference is given by $\Delta U=E_b$, and the work $\Delta W$ done 
upon the unrolling process can be written in terms of the interlayer pressure $p(d)$ \cite{SM} integrated over the interlayer separation as \cite{GREGORY1975}
\begin{eqnarray}
\Delta W & = & - A_\mathrm{o} \int\limits_d^\infty p(d=x) dx \\
         & = & -A_\mathrm{o} \epsilon_0 \epsilon \kappa \zeta^2 
 \left[1-\tanh\left(\kappa d/2\right)\right],
\label{work}
\end{eqnarray} 
where $\epsilon_0$ is the dielectric constant,
$\epsilon\approx 80$ is the dielectric permittivity for water solutions,
and $\kappa=\sqrt{2e^2\rho_\mathrm{ion}/(\epsilon_0 \epsilon k_B T)}$ is the reciprocal of the Debye length
for a 1:1 electrolyte with $\rho_\mathrm{ion}$ being the ion density.
The latter can be roughly estimated using pH and Avogadro number $N_A$ as 
$\rho_\mathrm{ion}=N_A\, 10^{-\mathrm{pH}}$~dm$^{-3}$ for pH $<7$ and $\rho_\mathrm{ion}=N_A\, 10^{\mathrm{pH}-14}$~dm$^{-3}$ for pH $>7$.
The interlayer disjoining force per unit area $p(d)$ can also be seen as an osmotic pressure \cite{israelachvili2015intermolecular}.
Scroll stability is determined by the sign of $\Delta {\cal H}$: the scrolls are stable if $\Delta {\cal H} >0$ 
({\it i.e.,} the unrolling process requires an external energy source) and unstable otherwise.

\begin{figure}[b]
  \includegraphics[width=\columnwidth]{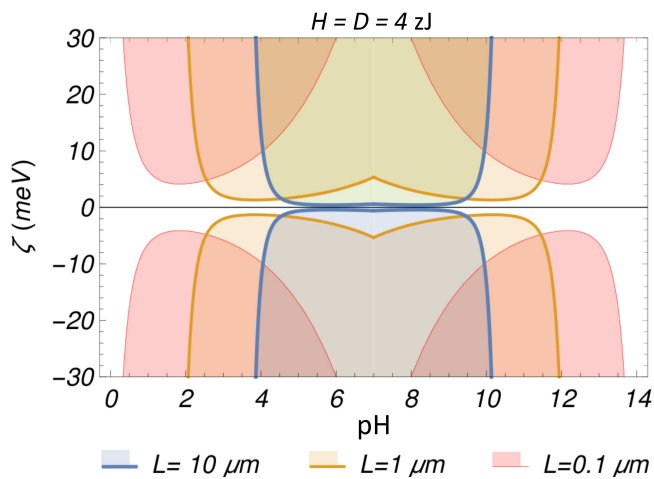}
  \caption{Phase diagram demonstrating the stability of scrolls immersed in an aqueous solution depending on
  the solution pH and $\zeta$ potential of functionalized graphene.
  The white regions correspond to $\Delta {\cal H} >0$ (the scrolls are stable for any size considered),
  and the color fillings indicate the instability regions for a given linear size $L$ shown in the legend.}
  \label{fig3}
\end{figure}

To investigate the sample-size effect, we set $A_\mathrm{o}\sim L^2$ in Eq.~(\ref{work}) and 
express $d$ through $L$ using Eqs. (\ref{archimed}) and (\ref{phi01}) as
$2\pi d \sim L/(1+\frac{1}{2}\sqrt{1+24\pi D/H})$.
Here, we have assumed $\varphi_{0,1} \gg 1$. We could also express $L$ in terms of $d$ but, in contrast to what we have done in Eq.~(\ref{energy}),
there is no way to get rid of both $L$ and $d$.
Also note that the electrostatic double-layer interactions are characterized by
{\em two} effective lengths: $1/\kappa$ and $e/\epsilon_0 \epsilon \zeta$.
The two-length dependence substantially expands the parameter space. In particular, 
the scroll stability can now be controlled externally through pH-dependent $\kappa$ and $\zeta$.

The interlayer distance $d$ is proportional to $L$ and depends on the ratio $D/H$.
In the limit $D/H \to \infty$ at a given $L$, the interlayer distance
formally vanishes ($d\to 0$). In the high-stiffness regime ($D/H \gg 1$),
the scroll shape resembles a rolled-up sheet of high-density office A4 paper
without crumples. The interlayer attraction is negligible for the macroscopic paper sheets 
(one has to hold a paper roll gently to keep it intact),
and the elastic force presses the layers together, resulting in a vanishing interlayer spacing.
Hence, the higher bending stiffness leads to a smaller interlayer distance.
In the case $D/H \sim 1$, one can imagine the scroll shape as a rolled-up sheet of cigarette paper,
which is much softer than office paper and forms looser scrolls. 
The scrolls collapse in the limit $D/H \ll 1$, in which the model does not apply.

Figure \ref{fig3} shows that the scrolls, once formed, are always stable at $\zeta\to 0$,
as there is no surface charge and hence no interlayer repulsion.
Away from the $\zeta = 0$ axis, the scroll stability is determined by the Debye length. 
The work $\Delta W$ vanishes at very low and very high ion densities ($\kappa\to 0$ and $\kappa\to\infty$),
always making $\Delta {\cal H}$ positive at those limits. That is the reason why the diagram demonstrates stability regions
at the neutral conditions (pH7) as well as in the strongly basic and strongly acidic solutions; see Fig.~\ref{fig3}.
The region of stability in pH-neutral solutions may, however, shrink to a narrow gap
if the repulsion force is strong (large $L$).
This is what one can see in Fig.~\ref{fig3} for $L=10$~$\mu$m. 
In the regions of very high ion concentrations (pH1 or pH13), the Debye length can become
shorter than the interlayer distance, and neighboring layers do not repel each other because of a strong screening.
It is probably not a technologically relevant regime because of the questionable material stability in such a
harsh environment, but it is instructive to consider this case for the sake of completeness.
After all, the Debye length can also be adjusted by adding a salt without making the solution
too acidic or too basic.


{\em Stability of fibers. ---} 
There is certainly a more conventional way to increase the stability of scrolls: entwining several scrolls
at once and hence forming a {\it fiber}. At the first sight, the fibers are more difficult to roll up
and keep stable because the bending stiffness of a stack increases with the number of layers $N$.
However, the interlayer attraction also gets stronger with an increasing $N$
because the interlayer distance is then reduced by a factor of $1/N$.
To quantify this effect, we consider the flakes being of the same size and write the following equation for each scroll component:
\begin{equation}
 r_n(\varphi) = \frac{\varphi_n +\varphi}{2\pi} d,\quad  \varphi_n= (n-1)\frac{2\pi}{N},\quad 1 \leq n \leq N.
 \label{fiber}
\end{equation}
Here, $n$ is the scroll index, and
the particular form of $\varphi_n$ is chosen to keep the layers equidistant; see Fig. \ref{fig4}.
Following the same recipe as before, we can derive the total fiber energy given by Eq.~(\ref{energy})
with $D\to ND$ and $H\to N^2 H$, {\it i.e.,} the interlayer attraction increases faster with $N$ than does the effective bending stiffness.
As a consequence, the fiber stability improves with the number of scroll components.

\begin{figure}
\includegraphics[width=\columnwidth]{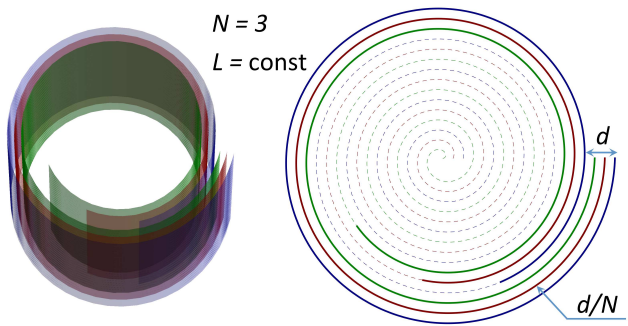}
\caption{Schematic image of a fiber made of three scrolls with $N=3$ in Eq. (\ref{fiber}). The size $L$ 
  can be calculated using the upper line of Eq. (\ref{archimed}), with $r(\varphi)$ given by Eq. (\ref{fiber}).
  The lower limit ($\varphi_1$) is adjusted to keep $L$ the same for all three scrolls.
  If the fiber components are made of the same material, then the interlayer distance is
  reduced by the factor $1/N$, making the fibers more stable than the individual scrolls.
  The dashed curves demonstrate the shift $\varphi_n$ from the coordinate origin of each scroll; see Eq. (\ref{fiber}). }
\label{fig4}
\end{figure}

{\em Outlook. ---} 
Because of its intrinsic universality, our model could be used as a compass 
for navigating in the space of external parameters determining the behavior of 2D scrolls.
Advanced functionalization of 2D materials can expand the parameter space even further, offering interesting regimes to explore.
One of the most obvious pathways is to functionalize the top and bottom of a pristine 2D flake by different chemicals \cite{GO-functionalization}.
This would result in a finite difference between $\zeta$ potentials on the inner and outer surfaces of the layer forming a scroll.
The difference qualitatively changes the electrostatic double-layer interactions, allowing for
the particular parameter combinations when the double-layer electrostatic repulsion switches to attraction \cite{GREGORY1975}.
This effect may either shrink the instability regions in the parameter space
or even lead to scroll collapse if the potential difference is too high.
The functionalization is therefore a powerful tool to change the geometry of 2D materials, 
one that should be used with care, however.


\acknowledgments
{\em Acknowledgements. ---}
We are grateful to Singapore NRF Medium-Sized Centre Programme for financial support.
M. T. is supported by the Director's Senior Research Fellowship at CA2DM (R-723-000-001-281).

\bibliography{scrolls.bib}

\begin{thebibliography}{47}%
\makeatletter
\providecommand \@ifxundefined [1]{%
 \@ifx{#1\undefined}
}%
\providecommand \@ifnum [1]{%
 \ifnum #1\expandafter \@firstoftwo
 \else \expandafter \@secondoftwo
 \fi
}%
\providecommand \@ifx [1]{%
 \ifx #1\expandafter \@firstoftwo
 \else \expandafter \@secondoftwo
 \fi
}%
\providecommand \natexlab [1]{#1}%
\providecommand \enquote  [1]{``#1''}%
\providecommand \bibnamefont  [1]{#1}%
\providecommand \bibfnamefont [1]{#1}%
\providecommand \citenamefont [1]{#1}%
\providecommand \href@noop [0]{\@secondoftwo}%
\providecommand \href [0]{\begingroup \@sanitize@url \@href}%
\providecommand \@href[1]{\@@startlink{#1}\@@href}%
\providecommand \@@href[1]{\endgroup#1\@@endlink}%
\providecommand \@sanitize@url [0]{\catcode `\\12\catcode `\$12\catcode
  `\&12\catcode `\#12\catcode `\^12\catcode `\_12\catcode `\%12\relax}%
\providecommand \@@startlink[1]{}%
\providecommand \@@endlink[0]{}%
\providecommand \url  [0]{\begingroup\@sanitize@url \@url }%
\providecommand \@url [1]{\endgroup\@href {#1}{\urlprefix }}%
\providecommand \urlprefix  [0]{URL }%
\providecommand \Eprint [0]{\href }%
\providecommand \doibase [0]{http://dx.doi.org/}%
\providecommand \selectlanguage [0]{\@gobble}%
\providecommand \bibinfo  [0]{\@secondoftwo}%
\providecommand \bibfield  [0]{\@secondoftwo}%
\providecommand \translation [1]{[#1]}%
\providecommand \BibitemOpen [0]{}%
\providecommand \bibitemStop [0]{}%
\providecommand \bibitemNoStop [0]{.\EOS\space}%
\providecommand \EOS [0]{\spacefactor3000\relax}%
\providecommand \BibitemShut  [1]{\csname bibitem#1\endcsname}%
\let\auto@bib@innerbib\@empty
\bibitem [{\citenamefont {Xu}\ \emph {et~al.}(2019)\citenamefont {Xu},
  \citenamefont {Wu}, \citenamefont {Huang},\ and\ \citenamefont
  {Mei}}]{REVIEW2019}%
  \BibitemOpen
  \bibfield  {author} {\bibinfo {author} {\bibfnamefont {C.}~\bibnamefont
  {Xu}}, \bibinfo {author} {\bibfnamefont {X.}~\bibnamefont {Wu}}, \bibinfo
  {author} {\bibfnamefont {G.}~\bibnamefont {Huang}}, \ and\ \bibinfo {author}
  {\bibfnamefont {Y.}~\bibnamefont {Mei}},\ }\href@noop {} {\bibfield
  {journal} {\bibinfo  {journal} {Advanced Materials Technologies}\ }\textbf
  {\bibinfo {volume} {4}},\ \bibinfo {pages} {1800486} (\bibinfo {year}
  {2019})}\BibitemShut {NoStop}%
\bibitem [{\citenamefont {Zhao}\ \emph {et~al.}(2021)\citenamefont {Zhao},
  \citenamefont {Wan}, \citenamefont {Liu}, \citenamefont {Xu}, \citenamefont
  {Yang}, \citenamefont {Shen}, \citenamefont {Zhang}, \citenamefont {Guo},
  \citenamefont {Qian}, \citenamefont {Li} \emph {et~al.}}]{nature2021}%
  \BibitemOpen
  \bibfield  {author} {\bibinfo {author} {\bibfnamefont {B.}~\bibnamefont
  {Zhao}}, \bibinfo {author} {\bibfnamefont {Z.}~\bibnamefont {Wan}}, \bibinfo
  {author} {\bibfnamefont {Y.}~\bibnamefont {Liu}}, \bibinfo {author}
  {\bibfnamefont {J.}~\bibnamefont {Xu}}, \bibinfo {author} {\bibfnamefont
  {X.}~\bibnamefont {Yang}}, \bibinfo {author} {\bibfnamefont {D.}~\bibnamefont
  {Shen}}, \bibinfo {author} {\bibfnamefont {Z.}~\bibnamefont {Zhang}},
  \bibinfo {author} {\bibfnamefont {C.}~\bibnamefont {Guo}}, \bibinfo {author}
  {\bibfnamefont {Q.}~\bibnamefont {Qian}}, \bibinfo {author} {\bibfnamefont
  {J.}~\bibnamefont {Li}},  \emph {et~al.},\ }\href@noop {} {\bibfield
  {journal} {\bibinfo  {journal} {Nature}\ }\textbf {\bibinfo {volume} {591}},\
  \bibinfo {pages} {385} (\bibinfo {year} {2021})}\BibitemShut {NoStop}%
\bibitem [{\citenamefont {Prinz}\ \emph {et~al.}(2000)\citenamefont {Prinz},
  \citenamefont {Seleznev}, \citenamefont {Gutakovsky}, \citenamefont
  {Chehovskiy}, \citenamefont {Preobrazhenskii}, \citenamefont {Putyato},\ and\
  \citenamefont {Gavrilova}}]{prinz2000}%
  \BibitemOpen
  \bibfield  {author} {\bibinfo {author} {\bibfnamefont {V.~Y.}\ \bibnamefont
  {Prinz}}, \bibinfo {author} {\bibfnamefont {V.~A.}\ \bibnamefont {Seleznev}},
  \bibinfo {author} {\bibfnamefont {A.~K.}\ \bibnamefont {Gutakovsky}},
  \bibinfo {author} {\bibfnamefont {A.~V.}\ \bibnamefont {Chehovskiy}},
  \bibinfo {author} {\bibfnamefont {V.~V.}\ \bibnamefont {Preobrazhenskii}},
  \bibinfo {author} {\bibfnamefont {M.~A.}\ \bibnamefont {Putyato}}, \ and\
  \bibinfo {author} {\bibfnamefont {T.~A.}\ \bibnamefont {Gavrilova}},\ }\href
  {\doibase https://doi.org/10.1016/S1386-9477(99)00249-0} {\bibfield
  {journal} {\bibinfo  {journal} {Physica E}\ }\textbf {\bibinfo {volume} {6}},\ \bibinfo {pages} {828 }
  (\bibinfo {year} {2000})}\BibitemShut {NoStop}%
\bibitem [{\citenamefont {Schmidt}\ and\ \citenamefont
  {Eberl}(2001)}]{schmidt2001nanotechnology}%
  \BibitemOpen
  \bibfield  {author} {\bibinfo {author} {\bibfnamefont {O.~G.}\ \bibnamefont
  {Schmidt}}\ and\ \bibinfo {author} {\bibfnamefont {K.}~\bibnamefont
  {Eberl}},\ }\href@noop {} {\bibfield  {journal} {\bibinfo  {journal}
  {Nature}\ }\textbf {\bibinfo {volume} {410}},\ \bibinfo {pages} {168}
  (\bibinfo {year} {2001})}\BibitemShut {NoStop}%
\bibitem [{\citenamefont {Shaji}\ \emph {et~al.}(2007)\citenamefont {Shaji},
  \citenamefont {Qin}, \citenamefont {Blick}, \citenamefont {Klein},
  \citenamefont {Deneke},\ and\ \citenamefont
  {Schmidt}}]{magnetotransport-2007schmidt}%
  \BibitemOpen
  \bibfield  {author} {\bibinfo {author} {\bibfnamefont {N.}~\bibnamefont
  {Shaji}}, \bibinfo {author} {\bibfnamefont {H.}~\bibnamefont {Qin}}, \bibinfo
  {author} {\bibfnamefont {R.~H.}\ \bibnamefont {Blick}}, \bibinfo {author}
  {\bibfnamefont {L.~J.}\ \bibnamefont {Klein}}, \bibinfo {author}
  {\bibfnamefont {C.}~\bibnamefont {Deneke}}, \ and\ \bibinfo {author}
  {\bibfnamefont {O.~G.}\ \bibnamefont {Schmidt}},\ }\href@noop {} {\bibfield
  {journal} {\bibinfo  {journal} {Applied Physics Letters}\ }\textbf {\bibinfo
  {volume} {90}},\ \bibinfo {pages} {042101} (\bibinfo {year}
  {2007})}\BibitemShut {NoStop}%
\bibitem [{\citenamefont {Schwaiger}\ \emph {et~al.}(2009)\citenamefont
  {Schwaiger}, \citenamefont {Br\"oll}, \citenamefont {Krohn}, \citenamefont
  {Stemmann}, \citenamefont {Heyn}, \citenamefont {Stark}, \citenamefont
  {Stickler}, \citenamefont {Heitmann},\ and\ \citenamefont
  {Mendach}}]{plasmon-2009mendach}%
  \BibitemOpen
  \bibfield  {author} {\bibinfo {author} {\bibfnamefont {S.}~\bibnamefont
  {Schwaiger}}, \bibinfo {author} {\bibfnamefont {M.}~\bibnamefont {Br\"oll}},
  \bibinfo {author} {\bibfnamefont {A.}~\bibnamefont {Krohn}}, \bibinfo
  {author} {\bibfnamefont {A.}~\bibnamefont {Stemmann}}, \bibinfo {author}
  {\bibfnamefont {C.}~\bibnamefont {Heyn}}, \bibinfo {author} {\bibfnamefont
  {Y.}~\bibnamefont {Stark}}, \bibinfo {author} {\bibfnamefont
  {D.}~\bibnamefont {Stickler}}, \bibinfo {author} {\bibfnamefont
  {D.}~\bibnamefont {Heitmann}}, \ and\ \bibinfo {author} {\bibfnamefont
  {S.}~\bibnamefont {Mendach}},\ }\href {\doibase
  10.1103/PhysRevLett.102.163903} {\bibfield  {journal} {\bibinfo  {journal}
  {Phys. Rev. Lett.}\ }\textbf {\bibinfo {volume} {102}},\ \bibinfo {pages}
  {163903} (\bibinfo {year} {2009})}\BibitemShut {NoStop}%
\bibitem [{\citenamefont {Trushin}\ and\ \citenamefont
  {Schliemann}(2007)}]{spin-2007trushin}%
  \BibitemOpen
  \bibfield  {author} {\bibinfo {author} {\bibfnamefont {M.}~\bibnamefont
  {Trushin}}\ and\ \bibinfo {author} {\bibfnamefont {J.}~\bibnamefont
  {Schliemann}},\ }\href {\doibase 10.1088/1367-2630/9/9/346} {\bibfield
  {journal} {\bibinfo  {journal} {New Journal of Physics}\ }\textbf {\bibinfo
  {volume} {9}},\ \bibinfo {pages} {346} (\bibinfo {year} {2007})}\BibitemShut
  {NoStop}%
\bibitem [{\citenamefont {Balhorn}\ \emph {et~al.}(2010)\citenamefont
  {Balhorn}, \citenamefont {Mansfeld}, \citenamefont {Krohn}, \citenamefont
  {Topp}, \citenamefont {Hansen}, \citenamefont {Heitmann},\ and\ \citenamefont
  {Mendach}}]{spin-2010mendach}%
  \BibitemOpen
  \bibfield  {author} {\bibinfo {author} {\bibfnamefont {F.}~\bibnamefont
  {Balhorn}}, \bibinfo {author} {\bibfnamefont {S.}~\bibnamefont {Mansfeld}},
  \bibinfo {author} {\bibfnamefont {A.}~\bibnamefont {Krohn}}, \bibinfo
  {author} {\bibfnamefont {J.}~\bibnamefont {Topp}}, \bibinfo {author}
  {\bibfnamefont {W.}~\bibnamefont {Hansen}}, \bibinfo {author} {\bibfnamefont
  {D.}~\bibnamefont {Heitmann}}, \ and\ \bibinfo {author} {\bibfnamefont
  {S.}~\bibnamefont {Mendach}},\ }\href {\doibase
  10.1103/PhysRevLett.104.037205} {\bibfield  {journal} {\bibinfo  {journal}
  {Phys. Rev. Lett.}\ }\textbf {\bibinfo {volume} {104}},\ \bibinfo {pages}
  {037205} (\bibinfo {year} {2010})}\BibitemShut {NoStop}%
\bibitem [{\citenamefont {Xie}\ \emph {et~al.}(2009)\citenamefont {Xie},
  \citenamefont {Ju}, \citenamefont {Feng}, \citenamefont {Sun}, \citenamefont
  {Zhou}, \citenamefont {Liu}, \citenamefont {Fan}, \citenamefont {Li},\ and\
  \citenamefont {Jiang}}]{graphene-scrolls-2009}%
  \BibitemOpen
  \bibfield  {author} {\bibinfo {author} {\bibfnamefont {X.}~\bibnamefont
  {Xie}}, \bibinfo {author} {\bibfnamefont {L.}~\bibnamefont {Ju}}, \bibinfo
  {author} {\bibfnamefont {X.}~\bibnamefont {Feng}}, \bibinfo {author}
  {\bibfnamefont {Y.}~\bibnamefont {Sun}}, \bibinfo {author} {\bibfnamefont
  {R.}~\bibnamefont {Zhou}}, \bibinfo {author} {\bibfnamefont {K.}~\bibnamefont
  {Liu}}, \bibinfo {author} {\bibfnamefont {S.}~\bibnamefont {Fan}}, \bibinfo
  {author} {\bibfnamefont {Q.}~\bibnamefont {Li}}, \ and\ \bibinfo {author}
  {\bibfnamefont {K.}~\bibnamefont {Jiang}},\ }\href {\doibase
  10.1021/nl900677y} {\bibfield  {journal} {\bibinfo  {journal} {Nano Letters}\
  }\textbf {\bibinfo {volume} {9}},\ \bibinfo {pages} {2565} (\bibinfo {year}
  {2009})}\BibitemShut {NoStop}%
\bibitem [{\citenamefont {Barcelos}\ \emph {et~al.}(2014)\citenamefont
  {Barcelos}, \citenamefont {Moura}, \citenamefont {Lacerda},\ and\
  \citenamefont {Malachias}}]{CVDgraphene-scrolls-2014}%
  \BibitemOpen
  \bibfield  {author} {\bibinfo {author} {\bibfnamefont {I.~D.}\ \bibnamefont
  {Barcelos}}, \bibinfo {author} {\bibfnamefont {L.~G.}\ \bibnamefont {Moura}},
  \bibinfo {author} {\bibfnamefont {R.~G.}\ \bibnamefont {Lacerda}}, \ and\
  \bibinfo {author} {\bibfnamefont {A.}~\bibnamefont {Malachias}},\ }\href
  {\doibase 10.1021/nl5012068} {\bibfield  {journal} {\bibinfo  {journal} {Nano
  Letters}\ }\textbf {\bibinfo {volume} {14}},\ \bibinfo {pages} {3919}
  (\bibinfo {year} {2014})}\BibitemShut {NoStop}%
\bibitem [{\citenamefont {Meng}\ \emph {et~al.}(2016)\citenamefont {Meng},
  \citenamefont {Wang}, \citenamefont {Li}, \citenamefont {Lu}, \citenamefont
  {Zhang}, \citenamefont {Yu}, \citenamefont {Chen}, \citenamefont {Du},
  \citenamefont {Liao}, \citenamefont {Zhao}, \citenamefont {Chen},
  \citenamefont {Zhu}, \citenamefont {Bai}, \citenamefont {Shi},\ and\
  \citenamefont {Zhang}}]{MoS2-scrolls-2016}%
  \BibitemOpen
  \bibfield  {author} {\bibinfo {author} {\bibfnamefont {J.}~\bibnamefont
  {Meng}}, \bibinfo {author} {\bibfnamefont {G.}~\bibnamefont {Wang}}, \bibinfo
  {author} {\bibfnamefont {X.}~\bibnamefont {Li}}, \bibinfo {author}
  {\bibfnamefont {X.}~\bibnamefont {Lu}}, \bibinfo {author} {\bibfnamefont
  {J.}~\bibnamefont {Zhang}}, \bibinfo {author} {\bibfnamefont
  {H.}~\bibnamefont {Yu}}, \bibinfo {author} {\bibfnamefont {W.}~\bibnamefont
  {Chen}}, \bibinfo {author} {\bibfnamefont {L.}~\bibnamefont {Du}}, \bibinfo
  {author} {\bibfnamefont {M.}~\bibnamefont {Liao}}, \bibinfo {author}
  {\bibfnamefont {J.}~\bibnamefont {Zhao}}, \bibinfo {author} {\bibfnamefont
  {P.}~\bibnamefont {Chen}}, \bibinfo {author} {\bibfnamefont {J.}~\bibnamefont
  {Zhu}}, \bibinfo {author} {\bibfnamefont {X.}~\bibnamefont {Bai}}, \bibinfo
  {author} {\bibfnamefont {D.}~\bibnamefont {Shi}}, \ and\ \bibinfo {author}
  {\bibfnamefont {G.}~\bibnamefont {Zhang}},\ }\href {\doibase
  10.1002/smll.201601413} {\bibfield  {journal} {\bibinfo  {journal} {Small}\
  }\textbf {\bibinfo {volume} {12}},\ \bibinfo {pages} {3770} (\bibinfo {year}
  {2016})}\BibitemShut {NoStop}%
\bibitem [{\citenamefont {Boukhvalov}\ and\ \citenamefont
  {Katsnelson}(2009)}]{graphene-functionalization-theory}%
  \BibitemOpen
  \bibfield  {author} {\bibinfo {author} {\bibfnamefont {D.~W.}\ \bibnamefont
  {Boukhvalov}}\ and\ \bibinfo {author} {\bibfnamefont {M.~I.}\ \bibnamefont
  {Katsnelson}},\ }\href {\doibase 10.1088/0953-8984/21/34/344205} {\bibfield
  {journal} {\bibinfo  {journal} {Journal of Physics: Condensed Matter}\
  }\textbf {\bibinfo {volume} {21}},\ \bibinfo {pages} {344205} (\bibinfo
  {year} {2009})}\BibitemShut {NoStop}%
\bibitem [{\citenamefont {Kuila}\ \emph {et~al.}(2012)\citenamefont {Kuila},
  \citenamefont {Bose}, \citenamefont {Mishra}, \citenamefont {Khanra},
  \citenamefont {Kim},\ and\ \citenamefont
  {Lee}}]{graphene-functionalization-review}%
  \BibitemOpen
  \bibfield  {author} {\bibinfo {author} {\bibfnamefont {T.}~\bibnamefont
  {Kuila}}, \bibinfo {author} {\bibfnamefont {S.}~\bibnamefont {Bose}},
  \bibinfo {author} {\bibfnamefont {A.~K.}\ \bibnamefont {Mishra}}, \bibinfo
  {author} {\bibfnamefont {P.}~\bibnamefont {Khanra}}, \bibinfo {author}
  {\bibfnamefont {N.~H.}\ \bibnamefont {Kim}}, \ and\ \bibinfo {author}
  {\bibfnamefont {J.~H.}\ \bibnamefont {Lee}},\ }\href@noop {} {\bibfield
  {journal} {\bibinfo  {journal} {Progress in Materials Science}\ }\textbf
  {\bibinfo {volume} {57}},\ \bibinfo {pages} {1061} (\bibinfo {year}
  {2012})}\BibitemShut {NoStop}%
\bibitem [{\citenamefont {Nguyen}\ \emph {et~al.}(2015)\citenamefont {Nguyen},
  \citenamefont {Carey}, \citenamefont {Ou}, \citenamefont {van Embden},
  \citenamefont {Gaspera}, \citenamefont {Chrimes}, \citenamefont {Spencer},
  \citenamefont {Zhuiykov}, \citenamefont {Kalantar-zadeh},\ and\ \citenamefont
  {Daeneke}}]{MoS2-functionalization}%
  \BibitemOpen
  \bibfield  {author} {\bibinfo {author} {\bibfnamefont {E.~P.}\ \bibnamefont
  {Nguyen}}, \bibinfo {author} {\bibfnamefont {B.~J.}\ \bibnamefont {Carey}},
  \bibinfo {author} {\bibfnamefont {J.~Z.}\ \bibnamefont {Ou}}, \bibinfo
  {author} {\bibfnamefont {J.}~\bibnamefont {van Embden}}, \bibinfo {author}
  {\bibfnamefont {E.~D.}\ \bibnamefont {Gaspera}}, \bibinfo {author}
  {\bibfnamefont {A.~F.}\ \bibnamefont {Chrimes}}, \bibinfo {author}
  {\bibfnamefont {M.~J.~S.}\ \bibnamefont {Spencer}}, \bibinfo {author}
  {\bibfnamefont {S.}~\bibnamefont {Zhuiykov}}, \bibinfo {author}
  {\bibfnamefont {K.}~\bibnamefont {Kalantar-zadeh}}, \ and\ \bibinfo {author}
  {\bibfnamefont {T.}~\bibnamefont {Daeneke}},\ }\href {\doibase
  10.1002/adma.201503163} {\bibfield  {journal} {\bibinfo  {journal} {Advanced
  Materials}\ }\textbf {\bibinfo {volume} {27}},\ \bibinfo {pages} {6225}
  (\bibinfo {year} {2015})}\BibitemShut {NoStop}%
\bibitem [{\citenamefont {Zhang}\ \emph {et~al.}(2019)\citenamefont {Zhang},
  \citenamefont {Guan}, \citenamefont {Ji}, \citenamefont {Liu}, \citenamefont
  {Jin},\ and\ \citenamefont {Xu}}]{GO-functionalization}%
  \BibitemOpen
  \bibfield  {author} {\bibinfo {author} {\bibfnamefont {M.}~\bibnamefont
  {Zhang}}, \bibinfo {author} {\bibfnamefont {K.}~\bibnamefont {Guan}},
  \bibinfo {author} {\bibfnamefont {Y.}~\bibnamefont {Ji}}, \bibinfo {author}
  {\bibfnamefont {G.}~\bibnamefont {Liu}}, \bibinfo {author} {\bibfnamefont
  {W.}~\bibnamefont {Jin}}, \ and\ \bibinfo {author} {\bibfnamefont
  {N.}~\bibnamefont {Xu}},\ }\href@noop {} {\bibfield  {journal} {\bibinfo
  {journal} {Nature Communications}\ }\textbf {\bibinfo {volume} {10}},\
  \bibinfo {pages} {1253} (\bibinfo {year} {2019})}\BibitemShut {NoStop}%
\bibitem [{\citenamefont {Berman}\ \emph {et~al.}(2015)\citenamefont {Berman},
  \citenamefont {Deshmukh}, \citenamefont {Sankaranarayanan}, \citenamefont
  {Erdemir},\ and\ \citenamefont {Sumant}}]{superlubricity-2015}%
  \BibitemOpen
  \bibfield  {author} {\bibinfo {author} {\bibfnamefont {D.}~\bibnamefont
  {Berman}}, \bibinfo {author} {\bibfnamefont {S.~A.}\ \bibnamefont
  {Deshmukh}}, \bibinfo {author} {\bibfnamefont {S.~K.}\ \bibnamefont
  {Sankaranarayanan}}, \bibinfo {author} {\bibfnamefont {A.}~\bibnamefont
  {Erdemir}}, \ and\ \bibinfo {author} {\bibfnamefont {A.~V.}\ \bibnamefont
  {Sumant}},\ }\href@noop {} {\bibfield  {journal} {\bibinfo  {journal}
  {Science}\ }\textbf {\bibinfo {volume} {348}},\ \bibinfo {pages} {1118}
  (\bibinfo {year} {2015})}\BibitemShut {NoStop}%
\bibitem [{\citenamefont {Zeng}\ \emph {et~al.}(2012)\citenamefont {Zeng},
  \citenamefont {Kuang}, \citenamefont {Liu}, \citenamefont {Liu},
  \citenamefont {Huang}, \citenamefont {Fu},\ and\ \citenamefont
  {Zhou}}]{supercapacitors-2012}%
  \BibitemOpen
  \bibfield  {author} {\bibinfo {author} {\bibfnamefont {F.}~\bibnamefont
  {Zeng}}, \bibinfo {author} {\bibfnamefont {Y.}~\bibnamefont {Kuang}},
  \bibinfo {author} {\bibfnamefont {G.}~\bibnamefont {Liu}}, \bibinfo {author}
  {\bibfnamefont {R.}~\bibnamefont {Liu}}, \bibinfo {author} {\bibfnamefont
  {Z.}~\bibnamefont {Huang}}, \bibinfo {author} {\bibfnamefont
  {C.}~\bibnamefont {Fu}}, \ and\ \bibinfo {author} {\bibfnamefont
  {H.}~\bibnamefont {Zhou}},\ }\href {\doibase 10.1039/C2NR30779K} {\bibfield
  {journal} {\bibinfo  {journal} {Nanoscale}\ }\textbf {\bibinfo {volume}
  {4}},\ \bibinfo {pages} {3997} (\bibinfo {year} {2012})}\BibitemShut
  {NoStop}%
\bibitem [{\citenamefont {Wang}\ \emph {et~al.}(2019)\citenamefont {Wang},
  \citenamefont {Bandari}, \citenamefont {Karnaushenko}, \citenamefont {Li},
  \citenamefont {Li}, \citenamefont {Zhang}, \citenamefont {Baunack},
  \citenamefont {Karnaushenko}, \citenamefont {Becker}, \citenamefont {Faghih},
  \citenamefont {Kang}, \citenamefont {Duan}, \citenamefont {Zhu},
  \citenamefont {Zhuang}, \citenamefont {Zhu}, \citenamefont {Feng},\ and\
  \citenamefont {Schmidt}}]{supercapacitors-poly-2019}%
  \BibitemOpen
  \bibfield  {author} {\bibinfo {author} {\bibfnamefont {J.}~\bibnamefont
  {Wang}}, \bibinfo {author} {\bibfnamefont {V.~K.}\ \bibnamefont {Bandari}},
  \bibinfo {author} {\bibfnamefont {D.}~\bibnamefont {Karnaushenko}}, \bibinfo
  {author} {\bibfnamefont {Y.}~\bibnamefont {Li}}, \bibinfo {author}
  {\bibfnamefont {F.}~\bibnamefont {Li}}, \bibinfo {author} {\bibfnamefont
  {P.}~\bibnamefont {Zhang}}, \bibinfo {author} {\bibfnamefont
  {S.}~\bibnamefont {Baunack}}, \bibinfo {author} {\bibfnamefont {D.~D.}\
  \bibnamefont {Karnaushenko}}, \bibinfo {author} {\bibfnamefont
  {C.}~\bibnamefont {Becker}}, \bibinfo {author} {\bibfnamefont
  {M.}~\bibnamefont {Faghih}}, \bibinfo {author} {\bibfnamefont
  {T.}~\bibnamefont {Kang}}, \bibinfo {author} {\bibfnamefont {S.}~\bibnamefont
  {Duan}}, \bibinfo {author} {\bibfnamefont {M.}~\bibnamefont {Zhu}}, \bibinfo
  {author} {\bibfnamefont {X.}~\bibnamefont {Zhuang}}, \bibinfo {author}
  {\bibfnamefont {F.}~\bibnamefont {Zhu}}, \bibinfo {author} {\bibfnamefont
  {X.}~\bibnamefont {Feng}}, \ and\ \bibinfo {author} {\bibfnamefont {O.~G.}\
  \bibnamefont {Schmidt}},\ }\href {\doibase 10.1021/acsnano.9b02917}
  {\bibfield  {journal} {\bibinfo  {journal} {ACS Nano}\ }\textbf {\bibinfo
  {volume} {13}},\ \bibinfo {pages} {8067} (\bibinfo {year}
  {2019})}\BibitemShut {NoStop}%
\bibitem [{\citenamefont {Costa}\ \emph {et~al.}(2021)\citenamefont {Costa},
  \citenamefont {Marangoni}, \citenamefont {Trushin}, \citenamefont {Carvalho},
  \citenamefont {Lim}, \citenamefont {Nguyen}, \citenamefont {Ng},
  \citenamefont {Zhao}, \citenamefont {Donato}, \citenamefont {Pennycook},
  \citenamefont {Sow}, \citenamefont {Novoselov},\ and\ \citenamefont
  {Castro~Neto}}]{ourAdvMat}%
  \BibitemOpen
  \bibfield  {author} {\bibinfo {author} {\bibfnamefont {M.~C.~F.}\
  \bibnamefont {Costa}}, \bibinfo {author} {\bibfnamefont {V.~S.}\ \bibnamefont
  {Marangoni}}, \bibinfo {author} {\bibfnamefont {M.}~\bibnamefont {Trushin}},
  \bibinfo {author} {\bibfnamefont {A.}~\bibnamefont {Carvalho}}, \bibinfo
  {author} {\bibfnamefont {S.~X.}\ \bibnamefont {Lim}}, \bibinfo {author}
  {\bibfnamefont {H.~T.~L.}\ \bibnamefont {Nguyen}}, \bibinfo {author}
  {\bibfnamefont {P.~R.}\ \bibnamefont {Ng}}, \bibinfo {author} {\bibfnamefont
  {X.}~\bibnamefont {Zhao}}, \bibinfo {author} {\bibfnamefont {R.~K.}\
  \bibnamefont {Donato}}, \bibinfo {author} {\bibfnamefont {S.~J.}\
  \bibnamefont {Pennycook}}, \bibinfo {author} {\bibfnamefont {C.~H.}\
  \bibnamefont {Sow}}, \bibinfo {author} {\bibfnamefont {K.~S.}\ \bibnamefont
  {Novoselov}}, \ and\ \bibinfo {author} {\bibfnamefont {A.~H.}\ \bibnamefont
  {Castro~Neto}},\ }\href {\doibase https://doi.org/10.1002/adma.202100442}
  {\bibfield  {journal} {\bibinfo  {journal} {Advanced Materials}\ }\textbf
  {\bibinfo {volume} {33}},\ \bibinfo {pages} {2100442} (\bibinfo {year}
  {2021})}\BibitemShut {NoStop}%
\bibitem [{\citenamefont {Coleman}(2009)}]{LPE2009coleman}%
  \BibitemOpen
  \bibfield  {author} {\bibinfo {author} {\bibfnamefont {J.~N.}\ \bibnamefont
  {Coleman}},\ }\href {\doibase 10.1002/adfm.200901640} {\bibfield  {journal}
  {\bibinfo  {journal} {Adv. Fun. Mater.}\ }\textbf {\bibinfo
  {volume} {19}},\ \bibinfo {pages} {3680} (\bibinfo {year}
  {2009})}\BibitemShut {NoStop}%
\bibitem [{\citenamefont {Sridhar}\ \emph {et~al.}(2010)\citenamefont
  {Sridhar}, \citenamefont {Jeon},\ and\ \citenamefont
  {Oh}}]{graphene-scrolls-chem}%
  \BibitemOpen
  \bibfield  {author} {\bibinfo {author} {\bibfnamefont {V.}~\bibnamefont
  {Sridhar}}, \bibinfo {author} {\bibfnamefont {J.-H.}\ \bibnamefont {Jeon}}, \
  and\ \bibinfo {author} {\bibfnamefont {I.-K.}\ \bibnamefont {Oh}},\ }\href
  {\doibase https://doi.org/10.1016/j.carbon.2010.04.034} {\bibfield  {journal}
  {\bibinfo  {journal} {Carbon}\ }\textbf {\bibinfo {volume} {48}},\ \bibinfo
  {pages} {2953 } (\bibinfo {year} {2010})}\BibitemShut {NoStop}%
\bibitem [{\citenamefont {Clower}\ \emph {et~al.}(2017)\citenamefont {Clower},
  \citenamefont {Groden},\ and\ \citenamefont {Wilson}}]{scrolls-ultrasonic}%
  \BibitemOpen
  \bibfield  {author} {\bibinfo {author} {\bibfnamefont {W.}~\bibnamefont
  {Clower}}, \bibinfo {author} {\bibfnamefont {N.}~\bibnamefont {Groden}}, \
  and\ \bibinfo {author} {\bibfnamefont {C.~G.}\ \bibnamefont {Wilson}},\
  }\href {\doibase https://doi.org/10.1016/j.nanoso.2017.09.005} {\bibfield
  {journal} {\bibinfo  {journal} {Nano-Structures \& Nano-Objects}\ }\textbf
  {\bibinfo {volume} {12}},\ \bibinfo {pages} {77 } (\bibinfo {year}
  {2017})}\BibitemShut {NoStop}%
\bibitem [{\citenamefont {Huang}\ \emph {et~al.}(2019)\citenamefont {Huang},
  \citenamefont {Huang}, \citenamefont {Liu}, \citenamefont {Zhou},
  \citenamefont {Ma}, \citenamefont {Wang}, \citenamefont {Qiu},\ and\
  \citenamefont {Bai}}]{GO-scrolls-lyophilization}%
  \BibitemOpen
  \bibfield  {author} {\bibinfo {author} {\bibfnamefont {X.}~\bibnamefont
  {Huang}}, \bibinfo {author} {\bibfnamefont {Z.}~\bibnamefont {Huang}},
  \bibinfo {author} {\bibfnamefont {Q.}~\bibnamefont {Liu}}, \bibinfo {author}
  {\bibfnamefont {A.}~\bibnamefont {Zhou}}, \bibinfo {author} {\bibfnamefont
  {Y.}~\bibnamefont {Ma}}, \bibinfo {author} {\bibfnamefont {J.}~\bibnamefont
  {Wang}}, \bibinfo {author} {\bibfnamefont {H.}~\bibnamefont {Qiu}}, \ and\
  \bibinfo {author} {\bibfnamefont {H.}~\bibnamefont {Bai}},\ }\href {\doibase
  10.1021/acsomega.9b00623} {\bibfield  {journal} {\bibinfo  {journal} {ACS
  Omega}\ }\textbf {\bibinfo {volume} {4}},\ \bibinfo {pages} {7420} (\bibinfo
  {year} {2019})}\BibitemShut {NoStop}%
\bibitem [{\citenamefont {Cendula}\ \emph {et~al.}(2011)\citenamefont
  {Cendula}, \citenamefont {Kiravittaya}, \citenamefont {M\"onch},
  \citenamefont {Schumann},\ and\ \citenamefont
  {Schmidt}}]{strained-roll-up-theory2010}%
  \BibitemOpen
  \bibfield  {author} {\bibinfo {author} {\bibfnamefont {P.}~\bibnamefont
  {Cendula}}, \bibinfo {author} {\bibfnamefont {S.}~\bibnamefont
  {Kiravittaya}}, \bibinfo {author} {\bibfnamefont {I.}~\bibnamefont
  {M\"onch}}, \bibinfo {author} {\bibfnamefont {J.}~\bibnamefont {Schumann}}, \
  and\ \bibinfo {author} {\bibfnamefont {O.~G.}\ \bibnamefont {Schmidt}},\
  }\href@noop {} {\bibfield  {journal} {\bibinfo  {journal} {Nano Letters}\
  }\textbf {\bibinfo {volume} {11}},\ \bibinfo {pages} {236} (\bibinfo {year}
  {2011})}\BibitemShut {NoStop}%
\bibitem [{\citenamefont {Alben}\ \emph {et~al.}(2011)\citenamefont {Alben},
  \citenamefont {Balakrisnan},\ and\ \citenamefont
  {Smela}}]{defect-roll-up-theory2011}%
  \BibitemOpen
  \bibfield  {author} {\bibinfo {author} {\bibfnamefont {S.}~\bibnamefont
  {Alben}}, \bibinfo {author} {\bibfnamefont {B.}~\bibnamefont {Balakrisnan}},
  \ and\ \bibinfo {author} {\bibfnamefont {E.}~\bibnamefont {Smela}},\ }\href
  {\doibase 10.1021/nl200473p} {\bibfield  {journal} {\bibinfo  {journal} {Nano
  Letters}\ }\textbf {\bibinfo {volume} {11}},\ \bibinfo {pages} {2280}
  (\bibinfo {year} {2011})}\BibitemShut {NoStop}%
\bibitem [{\citenamefont {Shi}\ \emph {et~al.}(2010)\citenamefont {Shi},
  \citenamefont {Pugno},\ and\ \citenamefont {Gao}}]{old-theory-review}%
  \BibitemOpen
  \bibfield  {author} {\bibinfo {author} {\bibfnamefont {X.}~\bibnamefont
  {Shi}}, \bibinfo {author} {\bibfnamefont {N.~M.}\ \bibnamefont {Pugno}}, \
  and\ \bibinfo {author} {\bibfnamefont {H.}~\bibnamefont {Gao}},\ }\href
  {\doibase 10.1016/S0894-9166(11)60002-5} {\bibfield  {journal} {\bibinfo
  {journal} {Acta Mechanica Solida Sinica}\ }\textbf {\bibinfo {volume} {23}},\
  \bibinfo {pages} {484} (\bibinfo {year} {2010})}\BibitemShut {NoStop}%
\bibitem [{\citenamefont {Yin}\ and\ \citenamefont
  {Shi}(2013)}]{Lennard-Jones-circles}%
  \BibitemOpen
  \bibfield  {author} {\bibinfo {author} {\bibfnamefont {Q.}~\bibnamefont
  {Yin}}\ and\ \bibinfo {author} {\bibfnamefont {X.}~\bibnamefont {Shi}},\
  }\href {\doibase 10.1039/C3NR00489A} {\bibfield  {journal} {\bibinfo
  {journal} {Nanoscale}\ }\textbf {\bibinfo {volume} {5}},\ \bibinfo {pages}
  {5450} (\bibinfo {year} {2013})}\BibitemShut {NoStop}%
\bibitem [{\citenamefont {Xia}\ \emph {et~al.}(2010)\citenamefont {Xia},
  \citenamefont {Xue}, \citenamefont {Xie}, \citenamefont {Chen}, \citenamefont
  {Lv}, \citenamefont {Besenbacher},\ and\ \citenamefont
  {Dong}}]{carbon-scrolls-mol-dyn}%
  \BibitemOpen
  \bibfield  {author} {\bibinfo {author} {\bibfnamefont {D.}~\bibnamefont
  {Xia}}, \bibinfo {author} {\bibfnamefont {Q.}~\bibnamefont {Xue}}, \bibinfo
  {author} {\bibfnamefont {J.}~\bibnamefont {Xie}}, \bibinfo {author}
  {\bibfnamefont {H.}~\bibnamefont {Chen}}, \bibinfo {author} {\bibfnamefont
  {C.}~\bibnamefont {Lv}}, \bibinfo {author} {\bibfnamefont {F.}~\bibnamefont
  {Besenbacher}}, \ and\ \bibinfo {author} {\bibfnamefont {M.}~\bibnamefont
  {Dong}},\ }\href {\doibase 10.1002/smll.201000646} {\bibfield  {journal}
  {\bibinfo  {journal} {Small}\ }\textbf {\bibinfo {volume} {6}},\ \bibinfo
  {pages} {2010} (\bibinfo {year} {2010})}\BibitemShut {NoStop}%
\bibitem [{\citenamefont {Braga}\ \emph {et~al.}(2004)\citenamefont {Braga},
  \citenamefont {Coluci}, \citenamefont {Legoas}, \citenamefont {Giro},
  \citenamefont {Galv\~{a}o},\ and\ \citenamefont
  {Baughman}}]{carbon-scrolls-mol-dyn2}%
  \BibitemOpen
  \bibfield  {author} {\bibinfo {author} {\bibfnamefont {S.~F.}\ \bibnamefont
  {Braga}}, \bibinfo {author} {\bibfnamefont {V.~R.}\ \bibnamefont {Coluci}},
  \bibinfo {author} {\bibfnamefont {S.~B.}\ \bibnamefont {Legoas}}, \bibinfo
  {author} {\bibfnamefont {R.}~\bibnamefont {Giro}}, \bibinfo {author}
  {\bibfnamefont {D.~S.}\ \bibnamefont {Galv\~{a}o}}, \ and\ \bibinfo {author}
  {\bibfnamefont {R.~H.}\ \bibnamefont {Baughman}},\ }\href {\doibase
  10.1021/nl0497272} {\bibfield  {journal} {\bibinfo  {journal} {Nano Letters}\
  }\textbf {\bibinfo {volume} {4}},\ \bibinfo {pages} {881} (\bibinfo {year}
  {2004})}\BibitemShut {NoStop}%
\bibitem [{\citenamefont {Tang}\ \emph {et~al.}(2018)\citenamefont {Tang},
  \citenamefont {Gao}, \citenamefont {Xiong}, \citenamefont {Dang},
  \citenamefont {Xu},\ and\ \citenamefont {Wang}}]{GO-roll-up-theory2018}%
  \BibitemOpen
  \bibfield  {author} {\bibinfo {author} {\bibfnamefont {B.}~\bibnamefont
  {Tang}}, \bibinfo {author} {\bibfnamefont {E.}~\bibnamefont {Gao}}, \bibinfo
  {author} {\bibfnamefont {Z.}~\bibnamefont {Xiong}}, \bibinfo {author}
  {\bibfnamefont {B.}~\bibnamefont {Dang}}, \bibinfo {author} {\bibfnamefont
  {Z.}~\bibnamefont {Xu}}, \ and\ \bibinfo {author} {\bibfnamefont
  {X.}~\bibnamefont {Wang}},\ }\href {\doibase 10.1021/acs.chemmater.8b02083}
  {\bibfield  {journal} {\bibinfo  {journal} {Chemistry of Materials}\ }\textbf
  {\bibinfo {volume} {30}},\ \bibinfo {pages} {5951} (\bibinfo {year}
  {2018})}\BibitemShut {NoStop}%
\bibitem [{\citenamefont {Poulin}\ \emph {et~al.}(2016)\citenamefont {Poulin},
  \citenamefont {Jalili}, \citenamefont {Neri}, \citenamefont {Nallet},
  \citenamefont {Divoux}, \citenamefont {Colin}, \citenamefont {Aboutalebi},
  \citenamefont {Wallace},\ and\ \citenamefont {Zakri}}]{GO-bending2016}%
  \BibitemOpen
  \bibfield  {author} {\bibinfo {author} {\bibfnamefont {P.}~\bibnamefont
  {Poulin}}, \bibinfo {author} {\bibfnamefont {R.}~\bibnamefont {Jalili}},
  \bibinfo {author} {\bibfnamefont {W.}~\bibnamefont {Neri}}, \bibinfo {author}
  {\bibfnamefont {F.}~\bibnamefont {Nallet}}, \bibinfo {author} {\bibfnamefont
  {T.}~\bibnamefont {Divoux}}, \bibinfo {author} {\bibfnamefont
  {A.}~\bibnamefont {Colin}}, \bibinfo {author} {\bibfnamefont {S.~H.}\
  \bibnamefont {Aboutalebi}}, \bibinfo {author} {\bibfnamefont
  {G.}~\bibnamefont {Wallace}}, \ and\ \bibinfo {author} {\bibfnamefont
  {C.}~\bibnamefont {Zakri}},\ }\href {\doibase 10.1073/pnas.1605121113}
  {\bibfield  {journal} {\bibinfo  {journal} {PNAS}\ }\textbf {\bibinfo {volume} {113}},\ \bibinfo {pages}
  {11088} (\bibinfo {year} {2016})}\BibitemShut {NoStop}%
\bibitem [{\citenamefont {Lu}\ \emph {et~al.}(2009)\citenamefont {Lu},
  \citenamefont {Arroyo},\ and\ \citenamefont {Huang}}]{graphene-bending-2009}%
  \BibitemOpen
  \bibfield  {author} {\bibinfo {author} {\bibfnamefont {Q.}~\bibnamefont
  {Lu}}, \bibinfo {author} {\bibfnamefont {M.}~\bibnamefont {Arroyo}}, \ and\
  \bibinfo {author} {\bibfnamefont {R.}~\bibnamefont {Huang}},\ }\href
  {\doibase 10.1088/0022-3727/42/10/102002} {\bibfield  {journal} {\bibinfo
  {journal} {Journal of Physics D: Applied Physics}\ }\textbf {\bibinfo
  {volume} {42}},\ \bibinfo {pages} {102002} (\bibinfo {year}
  {2009})}\BibitemShut {NoStop}%
\bibitem [{\citenamefont {Nelson}\ \emph {et~al.}(2004)\citenamefont {Nelson},
  \citenamefont {Piran},\ and\ \citenamefont
  {Weinberg}}]{nelson2004statistical}%
  \BibitemOpen
  \bibfield  {author} {\bibinfo {author} {\bibfnamefont {D.~R.}\ \bibnamefont
  {Nelson}}, \bibinfo {author} {\bibfnamefont {T.}~\bibnamefont {Piran}}, \
  and\ \bibinfo {author} {\bibfnamefont {S.}~\bibnamefont {Weinberg}},\
  }\href@noop {} {\emph {\bibinfo {title} {Statistical mechanics of membranes
  and surfaces}}}\ (\bibinfo  {publisher} {World Scientific},\ \bibinfo {year}
  {2004})\BibitemShut {NoStop}%
\bibitem [{\citenamefont {Castro~Neto}\ \emph {et~al.}(2009)\citenamefont
  {Castro~Neto}, \citenamefont {Guinea}, \citenamefont {Peres}, \citenamefont
  {Novoselov},\ and\ \citenamefont {Geim}}]{RevModPhys2009}%
  \BibitemOpen
  \bibfield  {author} {\bibinfo {author} {\bibfnamefont {A.~H.}\ \bibnamefont
  {Castro~Neto}}, \bibinfo {author} {\bibfnamefont {F.}~\bibnamefont {Guinea}},
  \bibinfo {author} {\bibfnamefont {N.~M.~R.}\ \bibnamefont {Peres}}, \bibinfo
  {author} {\bibfnamefont {K.~S.}\ \bibnamefont {Novoselov}}, \ and\ \bibinfo
  {author} {\bibfnamefont {A.~K.}\ \bibnamefont {Geim}},\ }\href {\doibase
  10.1103/RevModPhys.81.109} {\bibfield  {journal} {\bibinfo  {journal} {Rev.
  Mod. Phys.}\ }\textbf {\bibinfo {volume} {81}},\ \bibinfo {pages} {109}
  (\bibinfo {year} {2009})}\BibitemShut {NoStop}%
\bibitem [{\citenamefont
  {Israelachvili}(2015)}]{israelachvili2015intermolecular}%
  \BibitemOpen
  \bibfield  {author} {\bibinfo {author} {\bibfnamefont {J.~N.}\ \bibnamefont
  {Israelachvili}},\ }\href@noop {} {\emph {\bibinfo {title} {Intermolecular
  and surface forces}}}\ (\bibinfo  {publisher} {Academic press},\ \bibinfo
  {year} {2015})\BibitemShut {NoStop}%
\bibitem [{\citenamefont {French}(2000)}]{ceramics-hamaker-2000}%
  \BibitemOpen
  \bibfield  {author} {\bibinfo {author} {\bibfnamefont {R.~H.}\ \bibnamefont
  {French}},\ }\href@noop {} {\bibfield  {journal} {\bibinfo  {journal}
  {Journal of the American Ceramic Society}\ }\textbf {\bibinfo {volume}
  {83}},\ \bibinfo {pages} {2117} (\bibinfo {year} {2000})}\BibitemShut
  {NoStop}%
\bibitem [{\citenamefont {Dagastine}\ \emph {et~al.}(2002)\citenamefont
  {Dagastine}, \citenamefont {Prieve},\ and\ \citenamefont
  {White}}]{graphite-hamaker-2002}%
  \BibitemOpen
  \bibfield  {author} {\bibinfo {author} {\bibfnamefont {R.~R.}\ \bibnamefont
  {Dagastine}}, \bibinfo {author} {\bibfnamefont {D.~C.}\ \bibnamefont
  {Prieve}}, \ and\ \bibinfo {author} {\bibfnamefont {L.~R.}\ \bibnamefont
  {White}},\ }\href {\doibase https://doi.org/10.1006/jcis.2002.8239}
  {\bibfield  {journal} {\bibinfo  {journal} {Journal of Colloid and Interface
  Science}\ }\textbf {\bibinfo {volume} {249}},\ \bibinfo {pages} {78 }
  (\bibinfo {year} {2002})}\BibitemShut {NoStop}%
\bibitem [{\citenamefont {Chiou}\ \emph {et~al.}(2018)\citenamefont {Chiou},
  \citenamefont {Olukan}, \citenamefont {Almahri}, \citenamefont {Apostoleris},
  \citenamefont {Chiu}, \citenamefont {Lai}, \citenamefont {Lu}, \citenamefont
  {Santos}, \citenamefont {Almansouri},\ and\ \citenamefont
  {Chiesa}}]{graphene-hamaker-2018}%
  \BibitemOpen
  \bibfield  {author} {\bibinfo {author} {\bibfnamefont {Y.-C.}\ \bibnamefont
  {Chiou}}, \bibinfo {author} {\bibfnamefont {T.~A.}\ \bibnamefont {Olukan}},
  \bibinfo {author} {\bibfnamefont {M.~A.}\ \bibnamefont {Almahri}}, \bibinfo
  {author} {\bibfnamefont {H.}~\bibnamefont {Apostoleris}}, \bibinfo {author}
  {\bibfnamefont {C.~H.}\ \bibnamefont {Chiu}}, \bibinfo {author}
  {\bibfnamefont {C.-Y.}\ \bibnamefont {Lai}}, \bibinfo {author} {\bibfnamefont
  {J.-Y.}\ \bibnamefont {Lu}}, \bibinfo {author} {\bibfnamefont
  {S.}~\bibnamefont {Santos}}, \bibinfo {author} {\bibfnamefont
  {I.}~\bibnamefont {Almansouri}}, \ and\ \bibinfo {author} {\bibfnamefont
  {M.}~\bibnamefont {Chiesa}},\ }\href {\doibase 10.1021/acs.langmuir.8b02802}
  {\bibfield  {journal} {\bibinfo  {journal} {Langmuir}\ }\textbf {\bibinfo
  {volume} {34}},\ \bibinfo {pages} {12335} (\bibinfo {year}
  {2018})}\BibitemShut {NoStop}%
\bibitem [{\citenamefont {Rajter}\ \emph {et~al.}(2007)\citenamefont {Rajter},
  \citenamefont {French}, \citenamefont {Ching}, \citenamefont {Carter},\ and\
  \citenamefont {Chiang}}]{graphene-hamaker-comp-2007}%
  \BibitemOpen
  \bibfield  {author} {\bibinfo {author} {\bibfnamefont {R.~F.}\ \bibnamefont
  {Rajter}}, \bibinfo {author} {\bibfnamefont {R.~H.}\ \bibnamefont {French}},
  \bibinfo {author} {\bibfnamefont {W.~Y.}\ \bibnamefont {Ching}}, \bibinfo
  {author} {\bibfnamefont {W.~C.}\ \bibnamefont {Carter}}, \ and\ \bibinfo
  {author} {\bibfnamefont {Y.~M.}\ \bibnamefont {Chiang}},\ }\href {\doibase
  10.1063/1.2709576} {\bibfield  {journal} {\bibinfo  {journal} {Journal of
  Applied Physics}\ }\textbf {\bibinfo {volume} {101}},\ \bibinfo {pages}
  {054303} (\bibinfo {year} {2007})}\BibitemShut {NoStop}%
\bibitem [{\citenamefont {Feriancikova}\ and\ \citenamefont
  {Xu}(2012)}]{GO-hamaker-on-sand}%
  \BibitemOpen
  \bibfield  {author} {\bibinfo {author} {\bibfnamefont {L.}~\bibnamefont
  {Feriancikova}}\ and\ \bibinfo {author} {\bibfnamefont {S.}~\bibnamefont
  {Xu}},\ }\href {\doibase https://doi.org/10.1016/j.jhazmat.2012.07.041}
  {\bibfield  {journal} {\bibinfo  {journal} {Journal of Hazardous Materials}\
  }\textbf {\bibinfo {volume} {235-236}},\ \bibinfo {pages} {194} (\bibinfo
  {year} {2012})}\BibitemShut {NoStop}%
\bibitem [{\citenamefont {Gudarzi}(2016)}]{GO-Hamaker-comp-2016}%
  \BibitemOpen
  \bibfield  {author} {\bibinfo {author} {\bibfnamefont {M.~M.}\ \bibnamefont
  {Gudarzi}},\ }\href {\doibase 10.1021/acs.langmuir.6b01012} {\bibfield
  {journal} {\bibinfo  {journal} {Langmuir}\ }\textbf {\bibinfo {volume}
  {32}},\ \bibinfo {pages} {5058} (\bibinfo {year} {2016})}\BibitemShut
  {NoStop}%
\bibitem [{SM()}]{SM}%
  \BibitemOpen
  \href@noop {} {}\bibinfo {note} {See Supplemental Material for the
  electrokinetic theory basics.}\BibitemShut {Stop}%
\bibitem [{\citenamefont {Rice}\ and\ \citenamefont
  {Horne}(1981)}]{curvature1981}%
  \BibitemOpen
  \bibfield  {author} {\bibinfo {author} {\bibfnamefont {R.~E.}\ \bibnamefont
  {Rice}}\ and\ \bibinfo {author} {\bibfnamefont {F.~H.}\ \bibnamefont
  {Horne}},\ }\href {\doibase 10.1063/1.441936} {\bibfield  {journal} {\bibinfo
   {journal} {J. of Chem. Phys.}\ }\textbf {\bibinfo {volume}
  {75}},\ \bibinfo {pages} {5582} (\bibinfo {year} {1981})}\BibitemShut
  {NoStop}%
\bibitem [{\citenamefont {Rice}(1985)}]{curvature1985}%
  \BibitemOpen
  \bibfield  {author} {\bibinfo {author} {\bibfnamefont {R.~E.}\ \bibnamefont
  {Rice}},\ }\href {\doibase 10.1063/1.448826} {\bibfield  {journal} {\bibinfo
  {journal} {J. of Chem. Phys.}\ }\textbf {\bibinfo {volume}
  {82}},\ \bibinfo {pages} {4337} (\bibinfo {year} {1985})}\BibitemShut
  {NoStop}%
\bibitem [{\citenamefont {Shkel}\ \emph {et~al.}(2000)\citenamefont {Shkel},
  \citenamefont {Tsodikov},\ and\ \citenamefont {Record}}]{quasiplanar2000}%
  \BibitemOpen
  \bibfield  {author} {\bibinfo {author} {\bibfnamefont {I.~A.}\ \bibnamefont
  {Shkel}}, \bibinfo {author} {\bibfnamefont {O.~V.}\ \bibnamefont {Tsodikov}},
  \ and\ \bibinfo {author} {\bibfnamefont {M.~T.}\ \bibnamefont {Record}},\
  }\href@noop {} {\bibfield  {journal} {\bibinfo  {journal} {The Journal of
  Physical Chemistry B}\ }\textbf {\bibinfo {volume} {104}},\ \bibinfo {pages}
  {5161} (\bibinfo {year} {2000})}\BibitemShut {NoStop}%
\bibitem [{\citenamefont {Tuinier}(2003)}]{quasiplanar2003}%
  \BibitemOpen
  \bibfield  {author} {\bibinfo {author} {\bibfnamefont {R.}~\bibnamefont
  {Tuinier}},\ }\href {\doibase 10.1016/S0021-9797(02)00142-X} {\bibfield
  {journal} {\bibinfo  {journal} {J. of Colloid and Interface Sci.}\
  }\textbf {\bibinfo {volume} {258}},\ \bibinfo {pages} {45} (\bibinfo {year}
  {2003})}\BibitemShut {NoStop}%
\bibitem [{\citenamefont {Gregory}(1975)}]{GREGORY1975}%
  \BibitemOpen
  \bibfield  {author} {\bibinfo {author} {\bibfnamefont {J.}~\bibnamefont
  {Gregory}},\ }\href {\doibase https://doi.org/10.1016/0021-9797(75)90081-8}
  {\bibfield  {journal} {\bibinfo  {journal} {J. of Colloid and Interface Sci.}\ 
  }\textbf {\bibinfo {volume} {51}},\ \bibinfo {pages} {44 }
  (\bibinfo {year} {1975})}\BibitemShut {NoStop}%
\end{thebibliography}%

\begin{center}
{\Large SUPPLEMENTAL MATERIAL} 
\end{center}

Here, we derive the double-layer disjoining pressure $p(d)$ introduced in the main text. The Poisson-Boltzmann equation has
been solved in cylindrical coordinates long time ago, see Refs. \cite{curvature1981,curvature1985,quasiplanar2000,quasiplanar2003} in the main text,
and the present file is prepared for the sake of completeness.
The Poisson-Boltzmann equation reads
\begin{equation}
\label{P-B0}
 \nabla^2 \phi  = -\frac{e\rho}{\epsilon_0 \epsilon},
\end{equation}
where $\phi$ is the electrostatic potential, $\rho$ is the charge density given by the Boltzmann distribution,
$\epsilon_0$ is the dielectric constant, and $\epsilon\approx 80$ is the relative dielectric permittivity in water solutions.

We model the two neighboring layers in a scroll as two charged hollow cylinders.
We linearize the Poisson-Boltzmann equation assuming that the interlayer potential is lower than $k_B T$.
This approach is justified by the magnitude of the $\zeta$ potential typically being about $10$ meV.
We employ cylindrical coordinates and assume infinitely long cylinders with constant curvatures 
so that the resulting potential $\phi$ depends only on the radial coordinate, $r$. The equation reads
\begin{eqnarray}
  \frac{d^2 \phi_r}{dr^2} +\frac{1}{r}\frac{d \phi_r}{dr}& = & 
  -\frac{e\rho_\infty}{\epsilon_0 \epsilon}\left(\mathrm{e}^{\frac{-e\phi_r}{k_B T}} 
  -\mathrm{e}^{\frac{e\phi_r}{k_B T}}\right)\\
   & \simeq & \frac{2e\rho_\infty}{\epsilon_0 \epsilon} \frac{e\phi_r}{k_B T} \\
  & = & \kappa^2 \phi_r.
  \label{P-B}
 \end{eqnarray}
 Here, $k_B$ is the Boltzmann constant, $T=300$ K is the ambient temperature,  $\rho_\infty$ is the ionic concentration at $r\to\infty$, and
 \begin{equation}
 \kappa = \sqrt{\frac{2e^2 \rho_\infty}{\epsilon_0 \epsilon k_B T}}.
 \label{debye-eq}
\end{equation}
The quantity $1/\kappa$ is known as the Debye length. The general solution of equation (\ref{P-B}) reads
\begin{equation}
 \phi_r = A I_0(\kappa r) + B K_0(\kappa r),
\end{equation}
where $I_0$ and $K_0$ are the modified Bessel functions of the first and second kind.
To impose boundary conditions, we introduce two surface potentials $\phi_1 = \phi_{r=R_1}$ and $\phi_2 = \phi_{r=R_2}$.
We set $\phi_{1,2}=\zeta$ at the end of the day because both cylinders are made of the same material
and immersed into the same electrolyte and therefore characterized by the same $\zeta$ potential.
The general result for $\phi_1 \neq \phi_2$ reads 
\begin{eqnarray}
\nonumber \phi_r & =  & \frac{\phi_2 - \phi_1 \frac{K_0(\kappa R_2)}{K_0(\kappa R_1)}}{\frac{I_0(\kappa R_2)}{I_0(\kappa R_1)}
 -\frac{K_0(\kappa R_2)}{K_0(\kappa R_1)}}\frac{I_0(\kappa r)}{I_0(\kappa R_1)} \\
 && + \frac{\phi_1 \frac{I_0(\kappa R_2)}{I_0(\kappa R_1)} - \phi_2}{\frac{I_0(\kappa R_2)}{I_0(\kappa R_1)}
 -\frac{K_0(\kappa R_2)}{K_0(\kappa R_1)}}\frac{K_0(\kappa r)}{K_0(\kappa R_1)}.
 \label{solution}
\end{eqnarray}
One can prove by substituting equation (\ref{solution}) into equation (\ref{P-B}) that the former solves the latter.
Equation (\ref{solution}) is valid for any relation between $R_{1,2}$ and $\kappa$.

\begin{figure}
\begin{center}
\renewcommand{\thefigure}{S1}
\includegraphics[width=\columnwidth]{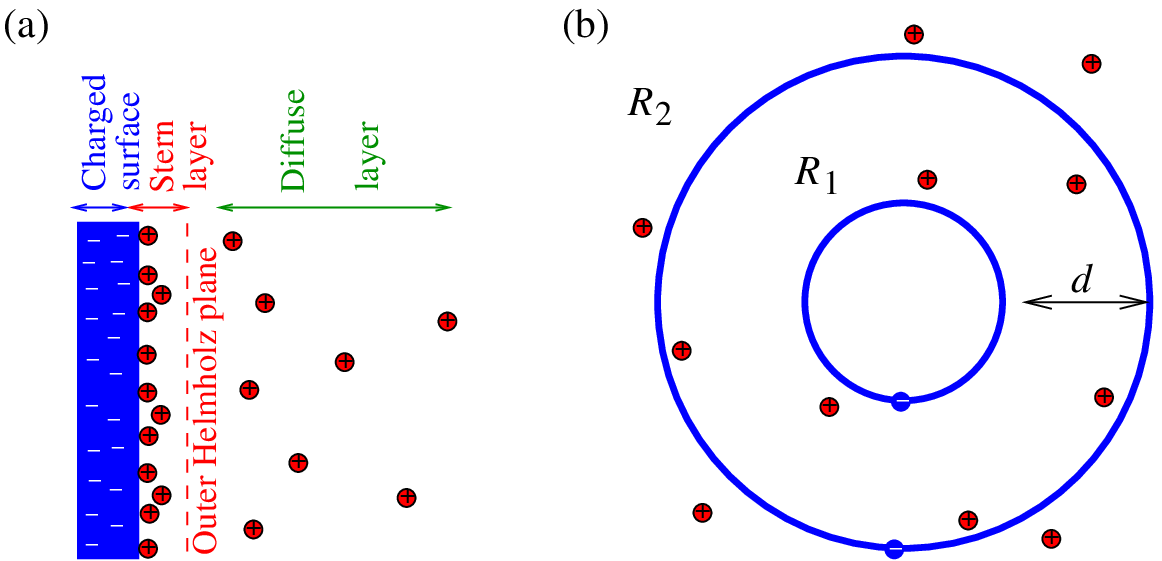}
 \caption{(a) Schematic representation of the counterion distribution at a charged interface. 
 The layer of immobile counterions near the charged surface is known as the Stern layer.
 The immobile ions do not contribute to the osmotic effect.
 The diffusive layer of mobile ions starts from the outer Helmholtz plane.
 The ions in the diffusive layer do not feel the true surface charge because of the screening Stern layer.
  The potential measured right next to the outer Helmholtz plane (in the slip plane) is the electrokinetic or  $\zeta$ potential.
  It is the $\zeta$ potential rather than the true surface potential that is responsible for the interlayer interactions.
 (b) The simplified model for electrostatic double-layer interactions between two neighboring layers.
 The interlayer distance, $d$, is translated into the difference between radii of two hollow cylinders enclosed in each other, $d=R_2-R_1$.
 In real scrolls, $d \ll R_{1,2}$ that justifies the quasiplanar approximation.
 The cylinders possess a negative surface charge attracting positively charged ions. The Stern layer is not shown in this panel.} 
 \label{approx}
\end{center}
\end{figure}

The double-layer force must be uniform throughout the interlayer gap at equilibrium, i.e.
independent of $r$, and it is also the pressure acting on the two layers given by
\begin{equation}
 p = \frac{\epsilon_0 \epsilon}{2} \kappa^2
 \left[ \left(\frac{1}{\kappa} \frac{d \phi_r}{dr}\right)^2
 -\phi_r^2 + 2\int\frac{dr}{r} \left(\frac{1}{\kappa} \frac{d \phi_r}{dr}\right)^2 \right].
 \label{result}
\end{equation}
Equation (\ref{result}) can be obtained by multiplying equation (\ref{P-B}) by ${d \phi_r}/{dr}$ and integrating over $r$. 
Making use of
\begin{eqnarray}
\nonumber \frac{1}{\kappa}\frac{d\phi_r}{dr} & = & \frac{\phi_2 - \phi_1 \frac{K_0(\kappa R_2)}{K_0(\kappa R_1)}}{\frac{I_0(\kappa R_2)}{I_0(\kappa R_1)}
 -\frac{K_0(\kappa R_2)}{K_0(\kappa R_1)}}\frac{I_1(\kappa r)}{I_0(\kappa R_1)}\\
&& - \frac{\phi_1 \frac{I_0(\kappa R_2)}{I_0(\kappa R_1)} - \phi_2}{\frac{I_0(\kappa R_2)}{I_0(\kappa R_1)}
 -\frac{K_0(\kappa R_2)}{K_0(\kappa R_1)}}\frac{K_1(\kappa r)}{K_0(\kappa R_1)},
 \label{solution2}
\end{eqnarray}
one can show by direct computation that the pressure {\em indeed} turns out to be independent of $r$, as expected.
It is determined by the interlayer separation and Debye length.
The explicit result can be expressed in terms of the Meijer functions.

The resulting expression is however too complicated to analyze. To make it simpler we assume that $\kappa R_{1,2} \gg 1$ and 
$R_2-R_1 \ll R_{1,2}$.
(Strictly speaking, the first assumption is invalid in neutral solutions where $\kappa$ is small, but
the pressure is negligible in this limit anyway.)
We use the following approximate relations $K_0(\kappa r)\simeq \sqrt{\pi/2\kappa r}e^{-\kappa r}$, $I_0(\kappa r)\simeq \sqrt{1/2\pi \kappa r}e^{\kappa r}$
and write
\begin{equation}
 \phi_r \sim \phi_1 \frac{\sinh\left[\kappa \left(R_2-r\right)\right]}{\sinh\left[\kappa \left(R_2-R_1\right)\right]}
 + \phi_2 \frac{\sinh\left[\kappa\left(r-R_1\right)\right])}{\sinh\left[\kappa \left(R_2-R_1\right)\right]}.
 \label{solution_approx}
\end{equation}
Making use of equation (\ref{solution_approx}), the interlayer pressure expression can be simplified strongly, and assuming 
symmetric boundary conditions $\phi_{1,2}=\zeta$ we arrive at the following expression
\begin{equation}
 p(d) = \epsilon_0\epsilon \kappa^2 \zeta^2  \frac{\cosh(\kappa d)-1}{\sinh^2(\kappa d)},
 \label{pressure}
\end{equation}
where $d=R_2-R_1$. Equation (\ref{pressure}) coincides with the expression derived in the planar limit
with the constant surface potentials, see Ref. [47] in the main text. The reason is our quasiplanar approximation applied above.
Equation (\ref{pressure}) is used in the main text to estimate the work done by the double-layer repulsive force for
disjoining the layers and unrolling the scroll.

\end{document}